\documentclass[twocolumn,prb,aps,showkeys,superscriptaddress,floatfix,amsmath,amssymb]{revtex4-1}
\usepackage[latin9]{inputenc}
\usepackage{verbatim}
\usepackage{amsmath}
\usepackage{amssymb}
\usepackage{graphicx}

\makeatletter

\providecommand{\tabularnewline}{\\}

\usepackage{dcolumn}% Align table columns on decimal point
\usepackage{bm}% bold math
\usepackage{tabularx}
\usepackage{color}
\usepackage{float}
\usepackage{amsfonts}
\usepackage[mathcal]{euscript}
\usepackage{array}
\usepackage{multirow}
\usepackage{natbib}
\usepackage[english]{babel}
\makeatother

\begin{document}
\title{Coulomb effects on topological band inversion in the moir\'e of WSe$_2$/BAs heterobilayer}
\author{Qizhong Zhu}
\email{zhuqizhong@gmail.com}
\affiliation{Department of Physics and Center of Theoretical and Computational
Physics, University of Hong Kong, Hong Kong, China}
\author{Qingjun Tong}
\affiliation{Department of Physics and Center of Theoretical and Computational
Physics, University of Hong Kong, Hong Kong, China}
\affiliation{School of Physics and Electronics, Hunan University, Changsha 410082,
China}
\author{Huazheng Sun}
\affiliation{School of Physics, Nankai University, Tianjin 300071, China}
\author{Yong Wang}
\affiliation{School of Physics, Nankai University, Tianjin 300071, China}
\author{Wang Yao}
\affiliation{Department of Physics and Center of Theoretical and Computational
Physics, University of Hong Kong, Hong Kong, China}

\begin{abstract}
Quantum spin Hall (QSH) insulator with large gap is highly desirable
for potential spintronics application. Here we realize electrically tunable
QSH insulator with large gap in van der Waals heterobilayer of monolayer
transition metal dichalcogenide (TMD) and hexagonal boron arsenide (BAs), in particular the WSe$_2$/BAs heterobilayer. When the type II band alignment gets inverted in an electric field, the hybridization by interlayer hopping between the spin-valley locked valence band edges in TMD and the BAs conduction band edges leads to a stacking-configuration dependent topological band inversion. In the non-interacting limit, the double spin degeneracy of BAs leaves an un-hybridized conduction band inside the gap, so the heterobilayer is a spin-valley locked metal instead of a QSH insulator. With the Coulomb interaction accounted in the double-layer geometry, the interaction with the hybridization induced electric dipole shifts this un-hybridized conduction band upwards in energy, giving rise to a sizable global QSH gap. Consequently, this heterobilayer provides a platform for engineering electrically tunable QSH insulator with sizable band gap. In the long-period moir\'e pattern with the spatial variation of local stacking-configurations, the competition between Coulomb interaction and interlayer hopping leads to superstructures of QSH insulators and excitonic insulators. 
\end{abstract}
\maketitle

\section{Introduction}

Quantum spin Hall (QSH) insulator \cite{hasan_textitcolloquium_2010,qi_topo_2011} featuring helical edge states immune from backscattering by nonmagnetic
impurities, is of great interest for the next-generation low-dissipation quantum electronic and spintronic devices. Material systems that can be electrically tuned between the normal insulating and the QSH insulating phases can be of particular interest, providing a mechanism of field effect transistors where the gate, instead of tuning the carrier density in a fixed conducting channel, simply creates/eliminates the topologically protected conducting channel in a bulk gap \cite{qian_quantum_2014,liu_switching_2015,tong_topological_2017}.
Concerning such desired electrical switchability, van der Waals heterostructures of two-dimensional (2D) materials can provide a unique approach towards QSH insulators, where the band alignment between the constituent layers can be most effectively tuned in the double gate geometry \cite{zhang_direct_2009}.

Theoretical study has predicted that heterobilayers of transition metal dichalcogenides (TMDs)
can be an electrically switchable QSH insulator \cite{tong_topological_2017}.
The TMD monolayers feature spin-valley locked massive Dirac cones located at the corners of hexagonal Brillouin zone \cite{coupled_xiao_2012,zeng_optical_2013,xu_spin_2014}. 
The interlayer hybridization of the spin-valley locked Dirac cones leads to either a topological band inversion or a trivial avoided band crossing, depending on the interlayer atomic registries.
This points to an intriguing platform for exploring QSH effect, with versatile tunability by interlayer bias and stacking registries. The sensitive dependence of the topological properties on the stacking configuration, arising from the valley nature of the band edges, further suggests unprecedented possibility for quantum engineering of topological superstructures in the moir\'e pattern of spatially varying local atomic configurations \cite{tong_topological_2017,hu_moire_2018}.
Theoretical study has further shown that the strong Coulomb interaction in the 2D geometry can turn the QSH insulator into a quantum anomalous Hall (QAH) insulator \cite{zhu_gate_2019}, which features chiral edge states that are completely lossless \cite{yu_quantized_2010,chang_experimental_2013,weng_quantum_2015}. 

In TMDs, the spin-orbit coupling is so strong that it gives rise to a spin-valley coupling of hundreds of meV. The size of the topological
gap from the interlayer hybridization is essentially determined by the strength of interlayer hopping between the two monolayer TMDs \cite{wang_interlayer_2017}. At the spin-valley locked band edges, the electronic states are predominantly from $d$-orbitals of the transition metal atoms \cite{voss_atomic_1999,lebegue_elec_2009,zhu_giant_2011,liu_three_2013,eugene_elec_2012,
chang_orbital_2013,capp_tight_2013}. The large separation of the active orbitals in the TMDs heterobilayers limits the interlayer hopping magnitude and hence the size of the topological gap. 
Moreover, the large band gap in the experimentally explored TMDs heterobilayers requires a large electric field to bring their type II alignment into the inverted regime. These are the two key issues lying between the intriguing concept of gate switchable topological superstructures in moir\'e and the experimental observation as well as practical applications. 

\begin{figure}[h]
\centering \includegraphics[width=8cm]{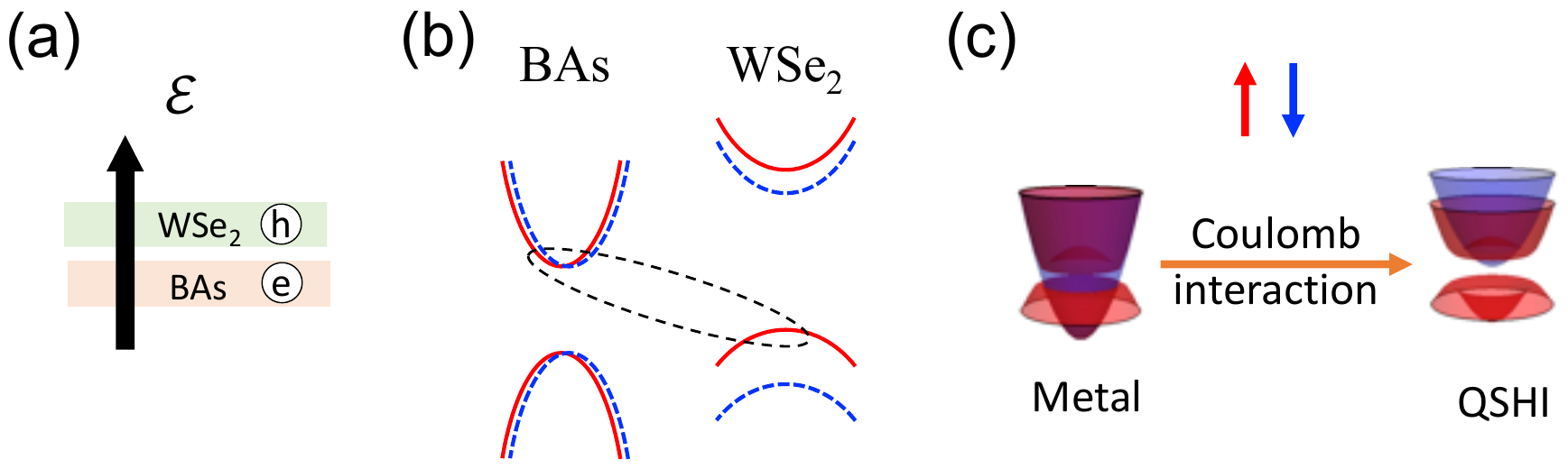} \caption{(Color online) (a) Schematic
of the heterobilayer. The electric field $\mathcal{E}$ tunes the relative band alignment of WSe$_2$ and hexagonal BAs.
(b) Type II band alignment of BAs and WSe$_{2}$ in $K$ valley without interlayer
bias. Spin doubly degenerate conduction band and spin-valley locked valence band
originate from the BAs and WSe$_{2}$ layers, respectively.
(c) Illustration of the effect of Coulomb interaction on the non-interacting band structure in $K$ valley.
In the non-interacting limit,
spin up (down) conduction and valence bands become hybridized in $K$
($-K$) valley, but the system is a metal because inside the hybridizaton gap,
there exists an un-hybridized conduction band from BAs. Taking into account the effect of Coulomb interaction,
the un-hybridized conduction bands are shifted upwards in energy due
to the interaction with the electric dipole associated with charge transfer between the two layers and hence a global band
gap is possible. The heterobilayer system can therefore realize a QSH insulator (QSHI). In (b) and (c), red (blue) band is spin up (down) polarized.}
\label{intro} 
\end{figure}

To address these outstanding issues, we consider here a heterobilayer formed by a monolayer hexagonal boron arsenide (BAs) and a TMD, with WSe$_2$ as an example, that features: (1) a much larger interlayer hopping because of the smaller interlayer separation between the band edge orbitals; (2) a small type II band gap that can be inverted by moderate gate bias \cite{Band_alignment}. 
Symmetry analysis shows that the conduction band edges of BAs monolayer are nearly spin-degenerate massive Dirac cones at the Brillouin zone corners. Their hybridization with the spin-valley locked valence band edges in TMD leads to topological band inversion depending on the stacking registry. In the non-interacting limit, the double spin degeneracy of BAs leaves an un-hybridized conduction band inside the hybridization gap, so the heterobilayer is a spin-valley locked metal. With Coulomb interaction accounted in the double-layer geometry, the interaction with the hybridization induced electric dipole shifts this un-hybridized conduction band upwards in energy, giving rise to a sizable global QSH gap, as illustrated in Fig. \ref{intro}. Using mean-field theory, we study the phase diagram in lattice-matched heterobilayers of TMD/BAs as function of dielectric constant, interlayer bias, and interlayer translation, which determine the relative importance of Coulomb interaction, the $s$-wave interlayer hopping, and the $p$-wave interlayer hopping in the band inversion. The competition of these factors in a long-period moir\'e pattern can lead to rich possibilities of electrically tunable superstructures consisting of large gap QSH insulators, excitonic insulators \cite{jerome_excitonic_1967} and exciton superfluids.

The paper is organized as follows. In Sec. II, we first study the
non-interacting properties of the TMD/BAs heterobilayer, taking WSe$_{2}$/BAs
heterobilayer as an example, based on the $\boldsymbol{k}\cdot\boldsymbol{p}$
model from symmetry analysis. Relevant parameters are obtained by
fitting the band structure from first-principles calculations. In
Sec. III, we study the Coulomb effect on the band inversion. Mean-field theory is used to self-consistently
take into account the effect of Coulomb interaction in the double-layer geometry. Interesting phases
like exciton superfluid and QSH/QAH insulators are found. The corresponding
results are presented in Sec. IV, including the phase diagrams for
AA-type and AB-type heterobilayers as functions of dielectric constant, interlayer bias, and interlayer translation, where the dependence on the last leads to gate tunable topological superstructures in long-period moir\'e. Sec. V discusses the relevant issues unexplored in detail in Sec. IV.

\section{Heterobilayer without Coulomb interaction}

Although yet to be synthesized,
2D hexagonal BAs monolayer is predicted to be stable in energetic,
dynamic, thermal, and mechanical properties, by extensive first-principles calculations \cite{sahin_monolayer_2009,xie_twod_2016,manoharan_exploring_2018}. Besides, the stable structure of hexagonal
BAs consists of B and As atoms located on the same plane \cite{sahin_monolayer_2009,xie_twod_2016,manoharan_exploring_2018}.
Here, we consider the heterobilayer formed by monolayer hexagonal BAs and WSe$_{2}$,
with lattice constants $a$
of 3.39 $\text{Å}$ and 3.325 $\text{Å}$, respectively. The small lattice constants
mismatch of $\sim2\%$ leads to a large moir\'e pattern with period $\sim50a$,
which can be further tuned by a relative strain and/or twisting. In
such a large moir\'e, one can regard each local region as commensurate
bilayer with identical lattice constants, while different local regions
differ by a relative in-plane displacement vector $\mathbf{R}$, which
defines a moir\'e supercell. The heterobilayer is then described
by commensurate bilayers parameterized by a vector $\mathbf{R}$.
In BAs, the valence band maximum and conduction band minimum are mainly
attributed to the $4p_{z}$ orbital of As and the $2p_{z}$ orbital
of B, respectively. Therefore, we define AA-type of stacking of WSe$_{2}$/BAs heterobilayer 
as B on W and As on Se atoms, as well as other configurations related
by interlayer translation; and AB-type of stacking as As on W and
B on Se atoms, as well as other configurations related by interlayer
translation. The in-plane displacement vector $\mathbf{R}$ is then
defined from W to B in AA stacking and from W to As in AB stacking.
To model the heterobilayer, in the following we first consider
the Hamiltonian of individual monolayer and then add a $\mathbf{R}$
dependent interlayer coupling.

The low energy physics of monolayer TMD is described by the massive
Dirac model. TMDs have a strong spin-orbital coupling originating
from the $d$-orbitals of the metal atoms. The valence band edge splitting
is about 0.15 eV for MoX$_{2}$ (X=S, Se) monolayer and about
0.45 eV for WX$_{2}$ (X=S, Se) monolayer. The conduction band edge splitting
is relatively smaller, about a few meV for MoS$_{2}$ and tens of
meV for WSe$_{2}$, WS$_{2}$ and MoSe$_{2}$. Around the $\pm K$ points
of Brillouin zone, the Hamiltonian of monolayer TMD is \cite{coupled_xiao_2012}
\begin{align}
H & =\frac{M_{1}}{2}\sigma_{z}+v_{1}\left(\tau k_{x}\sigma_{x}+k_{y}\sigma_{y}\right)\nonumber \\
 & +\tau s_{z}\left[\lambda_{c}\left(\sigma_{0}+\sigma_{z}\right)+\lambda_{v}\left(\sigma_{0}-\sigma_{z}\right)\right],
\end{align}
where $\tau=\pm1$ is the valley index and $\lambda_{c}$ $(\lambda_{v})$
is the spin-orbital coupling of conduction (valence) band. $\sigma$
denotes the Pauli matrices for the two basis functions $|d_{z^{2}}\rangle,\,\frac{1}{\sqrt{2}}\left(|d_{x^{2}-y^{2}}\rangle+i\tau|d_{xy}\rangle\right)$
and $s_{z}$ is the Pauli matrix for spin.

Monolayer BAs is a direct band gap semiconductor with band edges located
at $\pm K$ points \cite{sahin_monolayer_2009,xie_twod_2016,manoharan_exploring_2018}. Like graphene, we use two $p_{z}$ orbitals to
construct a tight-binding model 
\begin{equation}
H=\sum_{i}\mu_{i}c_{i}^{\dag}c_{i}+t\sum_{i,j}c_{i}^{\dag}c_{j}+h.c.,
\end{equation}
with only nearest-neighbor hopping considered. From fitting of
DFT band structure, we obtain the band gap $M_{2}=\mu_{\mathrm{B}}-\mu_{\mathrm{As}}=0.769$
eV, and hopping matrix element $t=1.43$ eV. Expanding around the $\pm K$ points in first order
of $\mathbf{k}$, we have 
\begin{equation}
H=\frac{M_{2}}{2}\sigma_{z}+v_{2}\left(\tau k_{x}\sigma_{x}+k_{y}\sigma_{y}\right),
\end{equation}
where the Fermi velocity $v_{2}=\sqrt{3}/2at=4.2$ eV$\text{Å}$ with the
lattice constant $a=3.39$~\AA. Neglecting the small spin-orbital
coupling, the band structure is spin degenerate.

In the long-period moir\'e pattern, each commensurate local can be described by
two massive Dirac models coupled by interlayer tunneling, dependent
on the relative translation $\mathbf{R}$. Neglecting the spin degree
of freedom, the bilayer Hamiltonian around the $K$ valley is \cite{tong_topological_2017}
\begin{align}
H_{\tau=1}(\mathbf{k},\mathbf{R}) & =\nonumber \\
 & \begin{bmatrix}-\frac{E_{g}}{2}+M_{1} & v_{1}k_{-\eta} & t_{cc}(\mathbf{R}) & t_{cv}(\mathbf{R})\\
v_{1}k_{+\eta} & -\frac{E_{g}}{2} & t_{vc}(\mathbf{R}) & t_{vv}(\mathbf{R})\\
t_{cc}^{*}(\mathbf{R}) & t_{vc}^{*}(\mathbf{R}) & \frac{E_{g}}{2} & v_{2}k_{-}\\
t_{cv}^{*}(\mathbf{R}) & t_{vv}^{*}(\mathbf{R}) & v_{2}k_{+} & \frac{E_{g}}{2}-M_{2}
\end{bmatrix},
\end{align}
where $\eta=1$ $(-1)$ for AA (AB) stacking, $M_{1}$ $(M_{2})$ is the band gap including spin-orbital splitting
in monolayer WSe$_{2}$ (BAs), and $E_{g}$ is the interlayer bias to
tune the relative alignment of the energy levels of different layers. $t_{ij}$ with $\{i,j\}=\{c,v\}$
are the hopping matrix elements between two layers. In addition, $k_{\pm}=k_{x}\pm ik_{y}$ and $k_{\pm\eta}=k_{x}\pm i\eta k_{y}$.

We are mainly interested in the regime of $E_{g}\sim0$, where the
four-band Hamiltonian can be approximately reduced to a two-band model composed
of relevant conduction band from BAs and valence band from WSe$_{2}$
with same spin polarization. The two-band Hamiltonian in such spin
species reads
\begin{gather}
H_{\tau}(\mathbf{k})\approx\begin{bmatrix}\frac{v_{2}^{2}}{M_{2}}k^{2}+\frac{E_{g}}{2} & t_{\tau\mathbf{k}}\\
t_{\tau\mathbf{k}}^{*} & -\frac{v_{1}^{2}}{M_{1}}k^{2}-\frac{E_{g}}{2}
\end{bmatrix},
\end{gather}
with $t_{\tau=1,\mathbf{k}}=t_{vc}^{*}+\frac{v_{2}}{M_{2}}t_{vv}^{*}k_{-}-\frac{v_{1}}{M_{1}}t_{cc}^{*}k_{-\eta}+\frac{v_{1}v_{2}}{M_{1}M_{2}}t_{cv}^{*}k_{-\eta}k_{-}$ and 
$t_{\tau=-1,\mathbf{k}}=t_{\tau=1,-\mathbf{k}}^{*}$.

When $E_{g}$ is tuned from positive to negative by the interlayer
bias, the conduction band from BAs and valence band from WSe$_{2}$
become inverted and cross at finite $\mathbf{k}$. The interlayer
coupling $t_{ij}$ opens a band gap around the crossing points. The
topology of the band gap thus opened is determined by the nature of interlayer hopping.
As the $t_{vc}$ term is $\mathbf{k}$ independent, the band
gap it opens is topologically trivial. On the other hand, the $t_{vv}$
and $t_{cc}$ terms are linear in $\mathbf{k}$ with different chirality,
and thus the band gaps opened by these two terms are topologically
nontrivial. The topological property of the $t_{cv}$ term depends
on the stacking. While the band gap is topologically trivial for AB
stacking, it is topologically nontrivial with chirality of 2 for AA
stacking. However, the band gap is relatively small as it is a
high-order process, and may be readily corrected in the presence of other
energy bands.

The interlayer hopping matrix element $t_{ij}$ depends on the interlayer atomic
registry, which is characterized by $\mathbf{R}$. By definition, $t_{\tau,ij}=\left\langle \Psi_{\tau,i}^{1}\left(\mathbf{R}\right)\right\vert H\left\vert \Psi_{\tau,j}^{2}\left(\mathbf{R}\right)\right\rangle $, where
$|\Psi_{\tau,i(j)}^{1(2)}\left(\mathbf{R}\right)\rangle$ denotes
the Bloch state at $\tau K$ point in the WSe$_{2}$ (BAs) layer.
For these three translation vectors $\mathbf{R}=0,\,\frac{\vec{a}_{1}+\vec{a}_{2}}{3},\,\frac{2(\vec{a}_{1}+\vec{a}_{2})}{3}$,
with $\vec{a}_{1}$, $\vec{a}_{2}$ being the primitive vectors spanning
the moir\'e supercell, the atomic configurations have three-fold
rotational symmetry. As $t_{\tau,ij}=\left\langle \hat{C}_{3}\Psi_{\tau,i}^{1}\right\vert H\left\vert \hat{C}_{3}\Psi_{\tau,j}^{2}\right\rangle =e^{i(\phi^{2}-\phi^{1})}\gamma_{\tau,m'}^{1*}\gamma_{\tau,m}^{2}t_{\tau,ij}$, with $\gamma_{\tau,m}^{1(2)}=e^{-i\frac{2m\pi}{3}}$ and $\phi^{1(2)}=0, \frac{2\pi}{3}, -\frac{2\pi}{3}$ for rotation centers at W (B) atom, Se (As) atom, hollow center respectively \cite{tong_topological_2017},
a non-zero interlayer hopping requires $e^{i(\phi^{2}-\phi^{1})}\gamma_{\tau,m'}^{1*}\gamma_{\tau,m}^{2}=1$.

The symmetry allowed interlayer hopping channels between conduction
and valence bands of the two layers at $K$ point are listed in the
following table. The magnitudes of these interlayer hoppings can be
evaluated from the first-principles calculations of band structures. Since
it is difficult to evaluate the interlayer coupling constants of a
heterobilayer directly, we estimate them from the results in the homobilayers
of BAs and WSe$_{2}$ using the formula $t_{ij}=\sqrt{t_{ij}^{\mathrm{BAs}}t_{ij}^{\mathrm{WSe_{2}}}}$.
The calculated coupling constants are several times larger than those in TMD bilayers, giving rise to much larger band gaps under the
same band inversion.

\begin{table}[htb]
\centering %
\noindent\begin{minipage}[t]{1\linewidth}%
\caption{Interlayer coupling matrix elements $t_{ij}$ (meV) at high symmetry
points of AA and AB stackings for $K$ valley. Different stacking registries are characterized
by different interlayer translations $R$ along the longer diagonal of the supercell.}
\begin{tabular}{c|c|c|c|c|c}
\hline
\hline 
 & $R/\sqrt{3}a$ & $t_{cc}$  & $t_{vv}$  & $t_{cv}$  & $t_{vc}$ \tabularnewline
\hline 
\hline
AA$_{0}$  & 0 & 33  & 0  & 0  & 0 \tabularnewline
\hline 
AA$_{1}$  & 1/3 & 0  & 98  & 0  & 0 \tabularnewline
\hline 
AA$_{2}$ & 2/3 & 0  & 0  & 106  & 106 \tabularnewline
\hline 
AB$_{0}$  & 0 & 0  & 0  & 81  & 0 \tabularnewline
\hline 
AB$_{1}$  & 1/3 & 0  & 0  & 0  & 81 \tabularnewline
\hline 
AB$_{2}$ & 2/3 & 46  & 113  & 0  & 0 \tabularnewline
\hline 
\hline
\end{tabular}\label{tab}%
\end{minipage}
\end{table}

At a general translation $\mathbf{R}$, the three-fold rotational
symmetry is broken. Under the two-center approximation, the $\mathbf{R}$
dependence of $t_{ij}$ can be formulated as 
\begin{align}
t_{cc}(\mathbf{R}) & \approx\frac{t_{cc}}{3}\left(e^{i\mathbf{K}\cdot\mathbf{R}}+e^{i\hat{C}_{3}\mathbf{K}\cdot\mathbf{R}}+e^{i\hat{C}_{3}^{2}\mathbf{K}\cdot\mathbf{R}}\right)\nonumber \\
t_{vv}(\mathbf{R}) & \approx\frac{t_{vv}}{3}\left(e^{i\mathbf{K}\cdot\mathbf{R}}+e^{i\left(\hat{C}_{3}\mathbf{K}\cdot\mathbf{R}-\frac{2\pi}{3}\right)}+e^{i\left(\hat{C}_{3}^{2}\mathbf{K}\cdot\mathbf{R}+\frac{2\pi}{3}\right)}\right)\nonumber \\
t_{cv}(\mathbf{R}) & \approx\frac{t_{cv}}{3}\left(e^{i\mathbf{K}\cdot\mathbf{R}}+e^{i\left(\hat{C}_{3}\mathbf{K}\cdot\mathbf{R}+\frac{2\pi}{3}\right)}+e^{i\left(\hat{C}_{3}^{2}\mathbf{K}\cdot\mathbf{R}-\frac{2\pi}{3}\right)}\right)\nonumber \\
t_{vc}(\mathbf{R}) & \approx\frac{t_{vc}}{3}\left(e^{i\mathbf{K}\cdot\mathbf{R}}+e^{i\left(\hat{C}_{3}\mathbf{K}\cdot\mathbf{R}+\frac{2\pi}{3}\right)}+e^{i\left(\hat{C}_{3}^{2}\mathbf{K}\cdot\mathbf{R}-\frac{2\pi}{3}\right)}\right),
\end{align}
for AA-type of WSe$_{2}$/BAs heterobilayer, and
\begin{align}
t_{cc}(\mathbf{R}) & \approx\frac{t_{cc}}{3}\left(e^{i\mathbf{K}\cdot\mathbf{R}}+e^{i\left(\hat{C}_{3}\mathbf{K}\cdot\mathbf{R}+\frac{2\pi}{3}\right)}+e^{i\left(\hat{C}_{3}^{2}\mathbf{K}\cdot\mathbf{R}-\frac{2\pi}{3}\right)}\right)\nonumber \\
t_{vv}(\mathbf{R}) & \approx\frac{t_{vv}}{3}\left(e^{i\mathbf{K}\cdot\mathbf{R}}+e^{i\left(\hat{C}_{3}\mathbf{K}\cdot\mathbf{R}+\frac{2\pi}{3}\right)}+e^{i\left(\hat{C}_{3}^{2}\mathbf{K}\cdot\mathbf{R}-\frac{2\pi}{3}\right)}\right)\nonumber \\
t_{cv}(\mathbf{R}) & \approx\frac{t_{cv}}{3}\left(e^{i\mathbf{K}\cdot\mathbf{R}}+e^{i\hat{C}_{3}\mathbf{K}\cdot\mathbf{R}}+e^{i\hat{C}_{3}^{2}\mathbf{K}\cdot\mathbf{R}}\right)\nonumber \\
t_{vc}(\mathbf{R}) & \approx\frac{t_{vc}}{3}\left(e^{i\mathbf{K}\cdot\mathbf{R}}+e^{i\left(\hat{C}_{3}\mathbf{K}\cdot\mathbf{R}-\frac{2\pi}{3}\right)}+e^{i\left(\hat{C}_{3}^{2}\mathbf{K}\cdot\mathbf{R}+\frac{2\pi}{3}\right)}\right),
\end{align}
for AB-type of WSe$_{2}$/BAs heterobilayer, where $\mathbf{K}$, $\hat{C}_{3}\mathbf{K}$ and $\hat{C}_{3}^{2}\mathbf{K}$ denote the wavevectors of the three corners of
the first Brillouin zone.

The conduction band from BAs is nearly spin degenerate, while the valence band
from WSe$_{2}$ is spin up (down) polarized at $K$ $\left(-K\right)$ valley
and separates energetically far away from band of the other spin species because
of the large spin-orbital splitting in the valence band. Taking into
account the spin degree of freedom, the system is described by a three-band
model,
\begin{align}
H_{\tau}(\mathbf{k},\mathbf{R}) & =\nonumber \\
 & \begin{bmatrix}\varepsilon_{\mathbf{k}c\bar{s}} & 0 & 0\\
0 & \varepsilon_{\mathbf{k}cs} & t_{\mathbf{k}}\left(\mathbf{R}\right)\\
0 & t_{\mathbf{k}}^{*}\left(\mathbf{R}\right) & \varepsilon_{\mathbf{k}vs}
\end{bmatrix},
\end{align}
with the basis $\left\{ \left\vert B,\bar{s}\right\rangle ,\,\left\vert B,s\right\rangle ,\,\left\vert W,s\right\rangle \right\} $. Here
$s=\uparrow\left(\downarrow\right)$ for $\tau=1\left(-1\right)$,
and $\bar{s}$ means the reverse of $s$.  Dispersions of conduction and valence bands are
$\varepsilon_{\mathbf{k}c}=\varepsilon_{\mathbf{k}cs(\bar{s})}=\frac{v_{2}^{2}}{M_{2}}k^{2}+\frac{E_{g}}{2}$
and $\varepsilon_{\mathbf{k}v}=\varepsilon_{\mathbf{k}vs}=-\frac{v_{1}^{2}}{M_{1}}k^{2}-\frac{E_{g}}{2}$,
respectively. The effective $\mathbf{R}$-dependent hopping matrix element is $t_{\mathbf{k}}\left(\mathbf{R}\right)=t_{vc}^{*}\left(\mathbf{R}\right)+\frac{v_{2}}{M_{2}}t_{vv}^{*}\left(\mathbf{R}\right)k_{-}-\frac{v_{1}}{M_{1}}t_{cc}^{*}\left(\mathbf{R}\right)k_{-\eta}$.

In the non-interacting limit, the interlayer hopping induces hybridization
of bands with same spin species, while the other conduction band is left
unaffected and lies inside the hybridization gap. The resulted phase is a metal with Fermi surfaces of opposite
spin polarization in different valleys (Fig. \ref{intro}(c)). We
will show in the following section that further taking into account
the effect of Coulomb interaction can shift the un-hybridized conduction
band upwards in energy and thus a global band gap is possible. In such a way, this heterobilayer
can realize a QSH insulator.

\section{The effect of Coulomb interaction}

The effect of Coulomb interaction can be taken into account by the
Hartree-Fock mean-field theory, which gives reasonable results in
treating other bilayer electron-hole systems \cite{zhu_exciton_1995,seradjeh_exciton_2009,pikulin_interplay_2014,budich_time_2014,wu_theory_2015,xue_time-reversal_2018}.
In mean-field theory, the ground state wave function of the three-band
system is  $|\Psi\rangle=\prod_{\tau\mathbf{k}}(u_{\tau\mathbf{k}}+v_{\tau\mathbf{k}s}\hat{a}_{\mathbf{k}s}^{\dagger}\hat{b}_{\mathbf{k}s}+v_{\tau\mathbf{k}\bar{s}}\hat{a}_{\mathbf{k}\bar{s}}^{\dagger}\hat{b}_{\mathbf{k}s})|0\rangle$,
where $s=\uparrow\left(\downarrow\right)$ for $\tau=1\left(-1\right)$,
$\bar{s}$ means the reverse of $s$, and $\hat{a}_{\mathbf{k}s}$
($\hat{b}_{\mathbf{k}s}$) is the annihilation operator of electron
with spin $s$ and momentum $\mathbf{k}$ in spin degenerate conduction
band (spin-valley locked valence band). $|0\rangle$ denotes the insulating state with
the valence bands of TMD completely filled. In the limit of vanishing interlayer
tunneling, it can be shown that the ground state of the three-band system
satisfies the condition $v_{\tau\mathbf{k}s}/v_{\tau\mathbf{k}\bar{s}}=\beta$,
with $\beta$ being a constant independent of $\mathbf{k}$ (see Appendix A).
Besides, the ground state energy is degenerate with respect to the
value of $\beta$, a manifestation of SU(2) symmetry of conduction
band electrons. Consequently, in terms of conduction band basis with spin orientation $\langle\hat{\mathbf{S}}\rangle_{\tau}$, the three-band wave function becomes
an effective two-band one,  $|\Psi\rangle=\prod_{\tau\mathbf{k}}(u_{\tau\mathbf{k}}+v_{\tau\mathbf{k}\langle\hat{\mathbf{S}}\rangle_{\tau}}\hat{a}_{\mathbf{k}\langle\hat{\mathbf{S}}\rangle_{\tau}}^{\dagger}\hat{b}_{\mathbf{k}s})|0\rangle$.
Here $\langle\hat{\mathbf{S}}\rangle_{\tau}$ is a function of $\beta$ but independent
of $\mathbf{k}$, defined as $\langle\hat{S_{i}}\rangle_{\tau}=\frac{\hbar}{2}(v_{\tau\mathbf{k}\uparrow}^{*},v_{\tau\mathbf{k}\downarrow}^{*})\sigma_{i}(v_{\tau\mathbf{k}\uparrow},v_{\tau\mathbf{k}\downarrow})^{T}/(|v_{\tau\mathbf{k}\uparrow}|^{2}+|v_{\tau\mathbf{k}\downarrow}|^{2})$, $\sigma_{i}$ being the Pauli matrices and $i=x,y,z$. The other
conduction band with opposite spin orientation is decoupled from the interaction induced hybridized bands, only shifted upwards
in energy due to the interaction with the hybridization induced electric dipole. Without loss of generality,
one can choose $v_{\tau\mathbf{k}\downarrow\left(\uparrow\right)}=0$
in $K$ ($-K$) valley, in which case $\langle\hat{\mathbf{S}}\rangle_{\tau}\sim\tau\boldsymbol{z}$.

The interlayer tunneling explicitly breaks the SU(2) symmetry of
conduction band electrons due to coupling with the spin-valley locked valence band. 
From an energetic point of view,
hybridization induced by both Coulomb interaction and interlayer tunneling should 
occur between conduction and valence bands with the same spin species to open the largest
possible gap, so the Coulomb interaction coupled conduction band should
have the same spin polarization with the spin-valley locked valence band. Therefore, only $v_{\tau\mathbf{k}\uparrow(\downarrow)}$
is nonzero in $K$ ($-K$) valley, and the other conduction band with
spin down (up) is still decoupled. This argument is also confirmed
by our numerical solution of a fully three-band self-consistent gap
equation with both Coulomb interaction and interlayer tunneling.

With the mean-field wave function  $|\Psi\rangle=\prod_{\tau\mathbf{k}}(u_{\tau\mathbf{k}}+v_{\tau\mathbf{k}s}\hat{a}_{\mathbf{k}s}^{\dagger}\hat{b}_{\mathbf{k}s})|0\rangle$,
the band coupling brought by both interlayer tunneling and Coulomb
interaction can be described by the mean-field Hamiltonian \cite{zhu_gate_2019}
in the original three-band basis, $\hat{H}_{\tau}=\sum_{\mathbf{k}}\left(\hat{a}_{\mathbf{k}\bar{s}}^{\dagger},\hat{a}_{\mathbf{k}s}^{\dagger},\hat{b}_{\mathbf{k}s}^{\dagger}\right)h_{\tau\mathbf{k}}\left(\hat{a}_{\mathbf{k}\bar{s}},\hat{a}_{\mathbf{k}s},\hat{b}_{\mathbf{k}s}\right)^{T}$,
and
\begin{align}
h_{\tau\mathbf{k}}=\begin{bmatrix}\xi_{\tau\mathbf{k}c\bar{s}} & 0 & 0\\
0 & \xi_{\tau\mathbf{k}cs} & -\Delta_{\tau\mathbf{k}}+t_{\tau\mathbf{k}}\\
0 & -\Delta_{\tau\mathbf{k}}^{*}+t_{\tau\mathbf{k}}^{*} & \xi_{\tau\mathbf{k}vs}
\end{bmatrix}.
\end{align}
The pairing gap function
$\Delta_{\tau\mathbf{k}}=\sum_{\mathbf{k}'}V\left(\mathbf{k}-\mathbf{k}'\right)u_{\tau\mathbf{k}'}^{*}v_{\tau\mathbf{k}'}$,
and $\xi_{\tau\mathbf{k}c(v)s}=\varepsilon_{\mathbf{k}c(v)}-\sum_{\mathbf{k}'}U\left(\mathbf{k}-\mathbf{k}'\right)|v_{\tau\mathbf{k}'}|^{2}+\frac{2\pi e^{2}d}{\epsilon}\sum_{\tau'\mathbf{k}'}|v_{\tau'\mathbf{k}'}|^{2}$,
$\xi_{\tau\mathbf{k}c\bar{s}}=\varepsilon_{\mathbf{k}c}+\frac{2\pi e^{2}d}{\epsilon}\sum_{\tau'\mathbf{k}'}|v_{\tau'\mathbf{k}'}|^{2}$.
Here $U(\mathbf{k})=2\pi e^{2}/(\epsilon k)$ and $V(\mathbf{k})=U(\mathbf{k})e^{-kd}$
are the intra- and inter-layer Coulomb interactions, respectively.
$\epsilon=\sqrt{\epsilon_{\parallel}\epsilon_{\perp}}$, where $\epsilon_{\parallel}$
($\epsilon_{\perp}$) is the intralayer (interlayer) dielectric constant.
$d=D\sqrt{\epsilon_{\parallel}/\epsilon_{\perp}}$, with $D$ being
the geometric interlayer distance. The last term in $\xi_{\tau\mathbf{k}c(v)s}$
is the classical charging energy of the bilayer as a parallel-plate
capacitor. The ground state wave function $|\Psi\rangle$ shall be
solved from the self-consistent gap equation, 
\begin{equation}
\Delta_{\tau\mathbf{k}}=\sum_{\mathbf{k}'}V\left(\mathbf{k}-\mathbf{k}'\right)\frac{\Delta_{\tau\mathbf{k}'}-t_{\tau\mathbf{k}'}}{2\sqrt{\zeta_{\tau\mathbf{k}'}^{2}+|\Delta_{\tau\mathbf{k}'}-t_{\tau\mathbf{k}'}|^{2}}},\label{gapeq}
\end{equation}
where $\zeta_{\tau\mathbf{k}}=\frac{\varepsilon_{\mathbf{k}c}-\varepsilon_{\mathbf{k}v}}{2}-\sum_{\mathbf{k}'}U\left(\mathbf{k}-\mathbf{k}'\right)|v_{\tau\mathbf{k}'}|^{2}+\frac{2\pi e^{2}d}{\epsilon}\sum_{\tau'\mathbf{k}'}|v_{\tau'\mathbf{k}'}|^{2}$.
The quasiparticle band dispersions are subsequently obtained as $E_{\tau\mathbf{k}cs}=\frac{\varepsilon_{\mathbf{k}c}+\varepsilon_{\mathbf{k}v}}{2}+\sqrt{\zeta_{\tau\mathbf{k}}^{2}+|\Delta_{\tau\mathbf{k}}-t_{\tau\mathbf{k}}|^{2}}$, $E_{\tau\mathbf{k}vs}=\frac{\varepsilon_{\mathbf{k}c}+\varepsilon_{\mathbf{k}v}}{2}-\sqrt{\zeta_{\tau\mathbf{k}}^{2}+|\Delta_{\tau\mathbf{k}}-t_{\tau\mathbf{k}}|^{2}}$, and
$E_{\tau\mathbf{k}c\bar{s}}=\varepsilon_{\mathbf{k}c}+\frac{2\pi e^{2}d}{\epsilon}\sum_{\tau\mathbf{k}}|v_{\tau\mathbf{k}}|^{2}$.
The global gap of the system is given by $\delta=\mathrm{min}\left\{\mathrm{min}\left(E_{\tau\mathbf{k}cs}-E_{\tau\mathbf{k}vs}\right),\mathrm{min}\left(E_{\tau\mathbf{k}c\bar{s}}-E_{\tau\mathbf{k}vs}\right)\right\}$. Both
the single particle dispersion and band hybridization gap are
renormalized by Coulomb interaction.

In the process of numerically solving the gap equation, there are
parameter regimes where multiple metastable solutions exist, in which case
the ground state is the one with lowest energy. 
Coulomb interaction favors a $s$-wave interlayer (electron-hole) pairing, 
and without interlayer tunneling $\left(t_{\tau\mathbf{k}}=0\right)$, the pairing gap function
$\Delta_{\tau\mathbf{k}}$ is a real function only dependent on $|\mathbf{k}|$. The interlayer
tunneling can have $s$-wave/$p$-wave component or their mixture,
depending on the stacking registries. 
The pairing gap function $\Delta_{\tau\mathbf{k}}$
plays the role of an effective interlayer tunneling, and its competition with
the bare interlayer tunneling leads to rich possibilities of topological and correlated phases. If interlayer tunneling
is $p$-wave dominant, the ground state can be topological phases
like QSH or QAH insulators, or exciton condensate with superfluid
properties. If interlayer tunneling is $s$-wave dominant, the ground
state will be a topologically trivial excitonic insulator (EI) without
superfluidity. The detailed phase diagrams are presented in the next
section.

\section{Phase diagram in different stackings}
\subsection{AA stacking}
\subsubsection{High symmetry points}

\begin{figure}[h]
\centering \includegraphics[width=8cm]{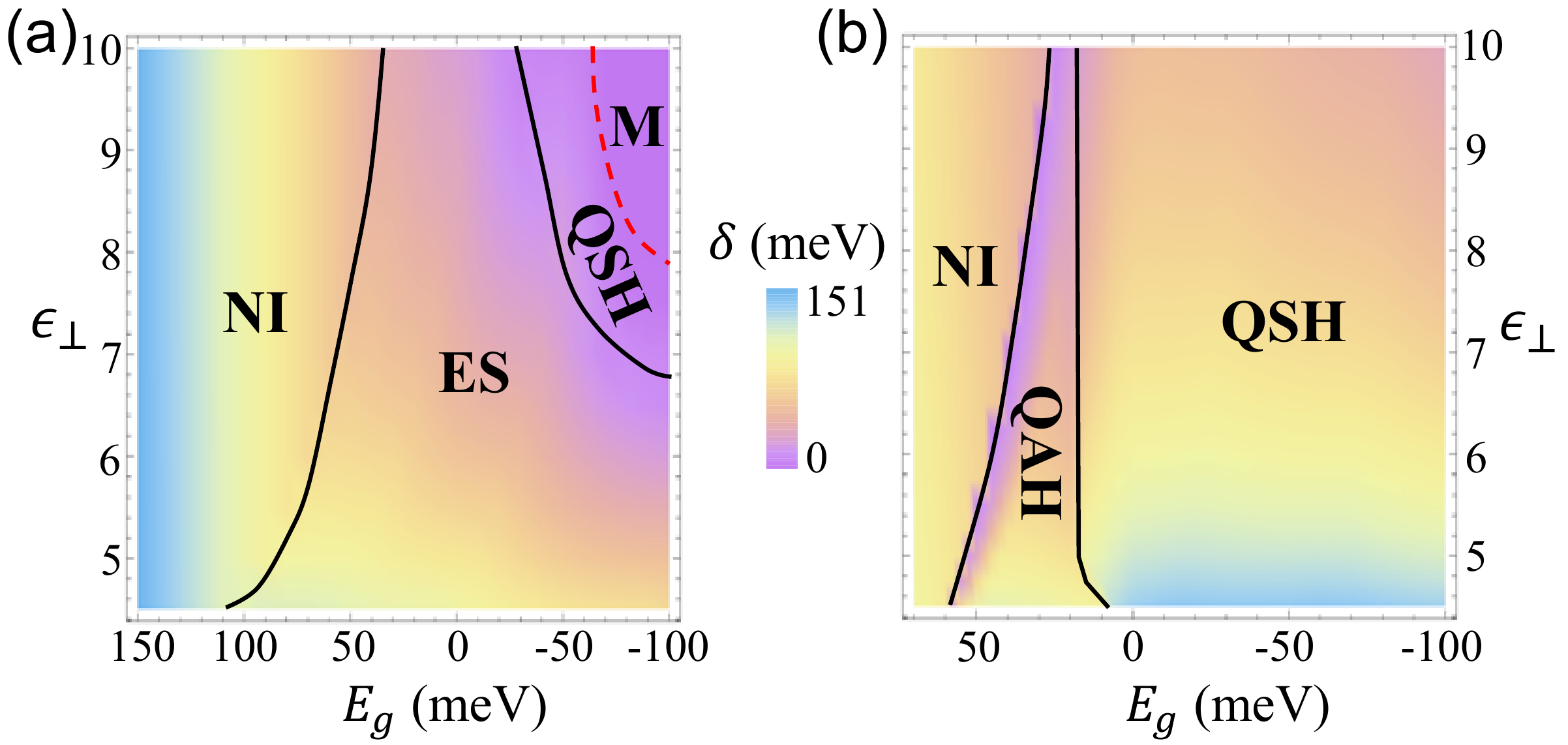} \caption{(Color online) Phase diagrams of AA stacking at two high symmetry points
AA$_{0}$ (a) and AA$_{1}$ (b), as functions of single particle
gap $E_{g}$ and interlayer dielectric constant $\epsilon_{\perp}$.
In the parameter regime shown here, the two phase diagrams both have
NI and QSH insulator phases. In phase diagram (a) with relatively weak interlayer tunneling, 
there are also ES
and metal phases, while in (b) with relatively strong interlayer tunneling, there is a QAH insulator phase. The red dashed line
marks the boundary of QSH insulator to metal phase transition, at which the global gap first
vanishes.}
\label{AA} 
\end{figure}

We first study the commensurate case, neglecting the small lattice mismatch in the
heterobilayer.
At the two high symmetry points AA$_{0}$
and AA$_{1}$ of AA stacking as defined in Table \ref{tab}, interlayer tunneling
can lead to topological phases upon band inversion. At AA$_{0}$,
only $t_{cc}$ is nonzero among the four interlayer hopping matrix
elements. So $t_{\tau\mathbf{k}}=-\frac{v_{1}}{M_{1}}t_{cc}^{*}\left(k_{x}-ik_{y}\right),$
with $v_{1}$ $(M_{1})$ being the Fermi velocity (band gap) in monolayer
WSe$_{2}$. Similarly, at AA$_{1}$, only $t_{vv}$ is nonzero. Then
$t_{\tau\mathbf{k}}=\frac{v_{2}}{M_{2}}t_{vv}^{*}\left(k_{x}-ik_{y}\right),$
with $v_{2}$ $(M_{2})$ being the Fermi velocity (band gap) in monolayer
BAs. For these two different stackings, the effective interlayer couplings
have different strengths. Coulomb effect on the band inversion and
phase diagrams in these two high symmetry points are shown in Fig. \ref{AA}.

In the numerical calculations, we fix $\epsilon_{\parallel}/\epsilon_{\perp}=1.6$
based on first-principles calculations of TMD bilayer \cite{kumar_tunable_2012},
$D=5.7 \textrm{Å}$ from first-principles calculations of WSe$_{2}$/BAs
heterobilayer, and parameters in the TMD massive Dirac model are obtained by fitting to first-principles band structures \cite{coupled_xiao_2012}.

\begin{figure}[h]
\centering \includegraphics[width=8cm]{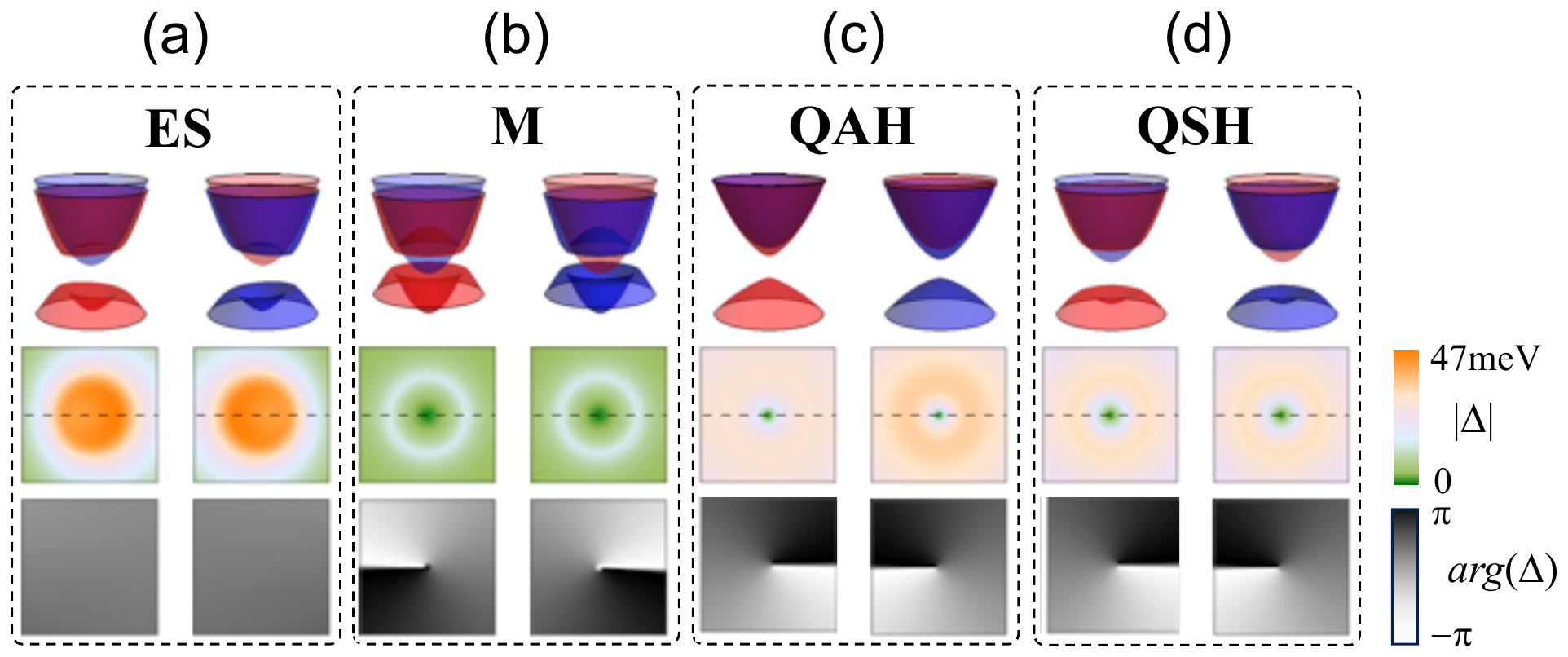} \caption{(Color online) Profiles
of the ES, metal, QAH and QSH phases ((a)-(d)) for a typical set of parameters, including quasiparticle
band structure, magnitude of order parameter $|\Delta|$ and phase
angle $arg(\Delta)$. The shown range of momentum space is $\left[-\frac{\pi}{10a},\,\frac{\pi}{10a}\right]\times\left[-\frac{\pi}{10a},\,\frac{\pi}{10a}\right]$.}
\label{pd} 
\end{figure}

In the parameter regime considered here, we find overall five phases.
Different phases are characterized by their different features of
pairing gap $\Delta_{\tau\mathbf{k}}$ (shown in Fig. \ref{pd}) and Hall conductance when the
Fermi level lies inside the global gap.
For these two stackings, their interlayer couplings both have $p$-wave component alone but
with different strengths,
while the Coulomb interaction favors a $s$-wave interlayer pairing.
The phase diagrams are resulted from their competition.

The phase diagrams in these two different stackings share some similarities.
At large single particle gap, the exciton density is vanishingly small,
and the ground state is a trivial normal insulator (NI). As the single
particle gap is reduced below the interlayer exciton binding energy,
the exciton density suddenly increases and the system can become a $s$-wave
exciton superfluid (ES) with minor effect brought by interlayer
tunneling, or a QAH insulator with spontaneous time reversal symmetry
breaking and nonzero Chern number. The former phase appears in the case of
AA$_{0}$ stacking (Fig. \ref{AA}(a)) with relatively weak interlayer
tunneling, while the latter appears in the case of AA$_{1}$ stacking (Fig. \ref{AA}(b))
where the interlayer tunneling is relatively strong. The spontaneous
time reversal symmetry breaking in the QAH phase can be understood in the light
of attractive intra-valley exciton exchange interaction as adopted by various
other researchers \cite{ciuti_role_1998,combescot_effects_2015,wu_theory_2015,zhu_gate_2019}.
With single particle gap further reduced, the system eventually enters
the QSH insulator phase, where interlayer tunneling dominates and
Coulomb interaction renders a global gap by shifting the un-hybridized
conduction band of BAs layer upwards in energy through the hybridization induced electric dipole. For the case of AA$_{0}$ stacking, there is also
a parameter regime where the un-hybridized conduction band lies inside
the hybridized band gap, resulting in a metal phase (M phase in Fig.
\ref{AA}(a)).

In the QSH phase, the overall layer hybridized band gap is strongly enhanced
by Coulomb interaction compared with non-interacting limit, and sizable
band gap along with electrical switchability of topological phases points to potential applications of this heterobilayer as dissipationless spintronic devices.

\subsubsection{General translation}

\begin{figure}[h]
\centering \includegraphics[width=8cm]{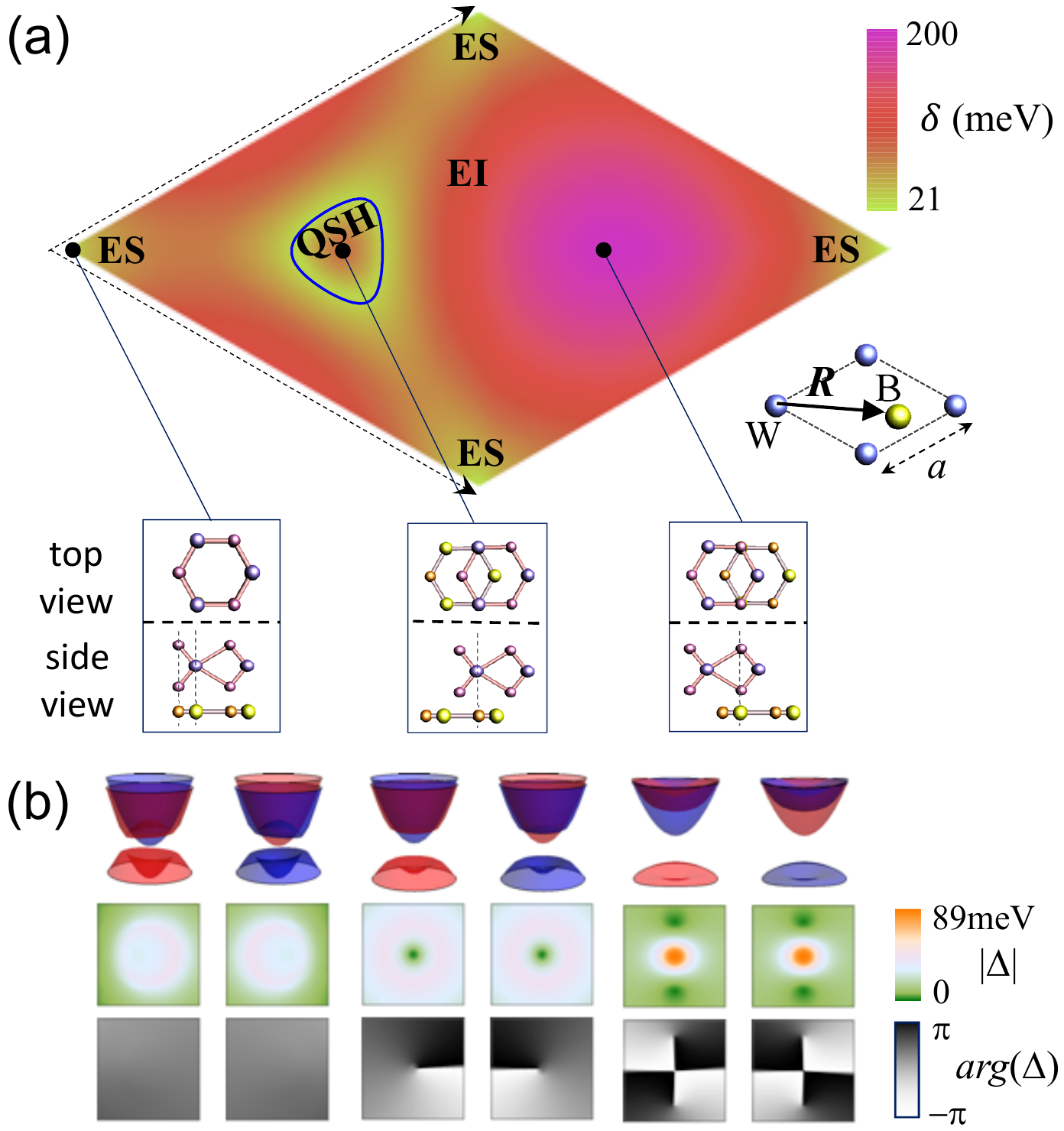} \caption{(Color online) (a) Phase diagram of AA stacking in a moir\'e supercell. There are three phases, namely, ES, EI and
QSH insulator. The QSH/EI phase boundary is marked by blue line. The nonzero gap at the
boundary is due to finite discretization interval of $\mathbf{k}$ in numerical calculations.
$\epsilon_{\perp}=6$ and $E_g=-100$ meV.
(b) Phases at three high symmetry locals along with top and side views
of these three stackings. The shown range of momentum space is $\left[-\frac{\pi}{10a},\,\frac{\pi}{10a}\right]\times\left[-\frac{\pi}{10a},\,\frac{\pi}{10a}\right]$, except in the $|\Delta|$
and $arg(\Delta)$ plots of the last stacking registry, where the range is
$\left[-\frac{\pi}{2a},\,\frac{\pi}{2a}\right]\times\left[-\frac{\pi}{2a},\,\frac{\pi}{2a}\right]$,
for a better illustration of the double vortices structure in $\Delta$. }
\label{AAr} 
\end{figure}

The small lattice mismatch between WSe$_{2}$ and BAs leads to moir\'e
superlattice in their heterobilayer. For long-period moir\'e pattern,
different regions in the moir\'e supercell are equivalent to commensurate
bilayers with different interlayer translations. Modeling the moir\'e
supercell by spatially varying interlayer tunneling, the phase diagram
in the supercell can be subsequently obtained.

For AA stacking, the phase diagram is shown in Fig. \ref{AAr}, where we
fix $\epsilon_{\perp}=6$ and the band inversion as $E_g=-100$
meV. In this parameter regime, only one high symmetry local AA$_{1}$ can support topological phases.

For regions close enough to the high symmetry local, the weight
of $s$-wave component is relatively small and the QSH phase is still
preserved.
Away from the high symmetry local AA$_{1}$, the strength of $s$-wave
tunneling gradually increases, tending to turn the
QSH phase into topologically trivial EI phase
with the phase of pairing gap completely pinned and superfluidity
lost. At the boundary of QSH to EI phase transition,
band gap closes as in non-interacting limit (The finite band gap at
the boundary here is due to the finite interval in numerical discretization
 of $\mathbf{k}$ and decreases for denser $\mathbf{k}$-point meshes). 
 Note that ES phases are also present for regions close to the
 high symmetry local AA$_{0}$ due to vanishing $s$-wave tunneling there,
 as further discussed in detail in Sec. V.

\subsection{AB stacking}
\subsubsection{High symmetry points}

\begin{figure}[h]
\centering \includegraphics[width=8cm]{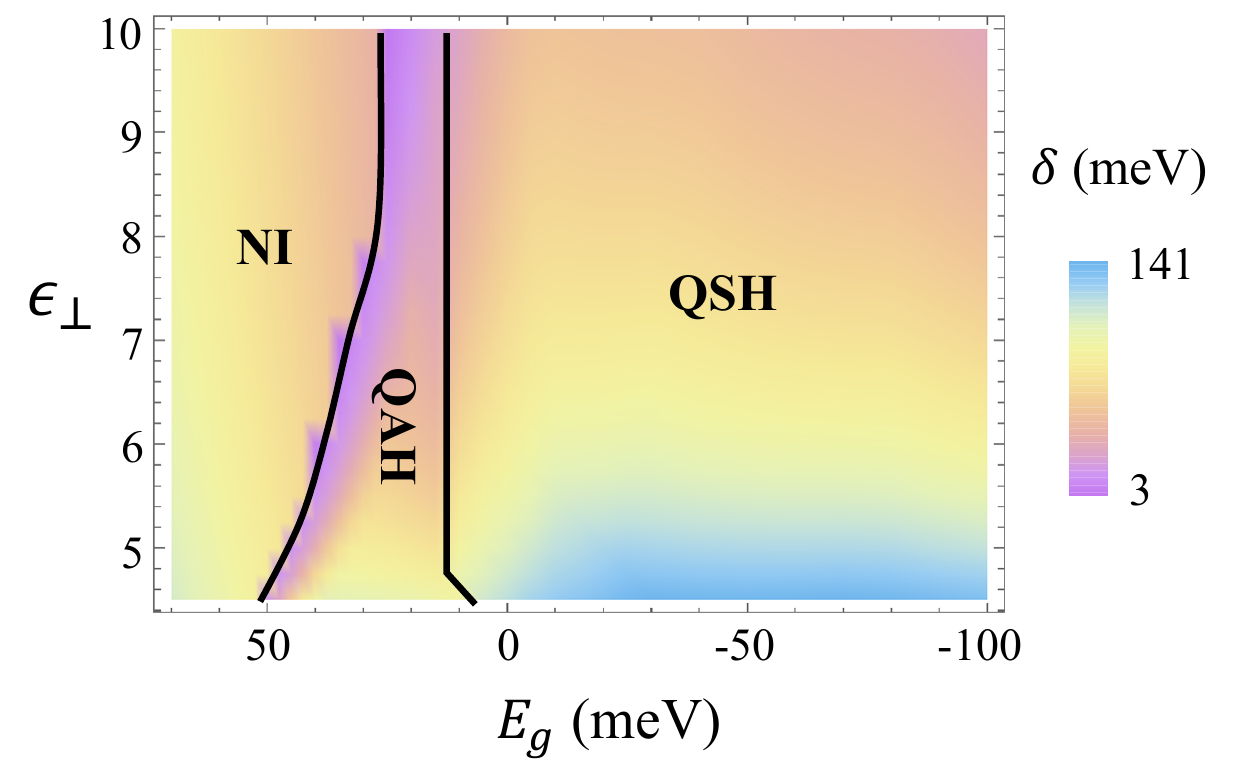} \caption{(Color online) Phase diagram of AB stacking at high symmetry point AB$_2$,
as a function of single particle gap $E_{g}$ and interlayer dielectric
constant $\epsilon_{\perp}$. The three phases are similar to those
in Fig. \ref{AA}, but are slightly anisotropic in $|\Delta|$.}
\label{AB} 
\end{figure}

Next, we consider the AB stacking case in lattice-matched heterobilayer. Only one high symmetry point
AB$_{2}$ can support topological phases. At AB$_{2}$, $t_{\tau\mathbf{k}}=\frac{v_{2}}{M_{2}}t_{vv}^{*}\left(k_{x}-ik_{y}\right)-\frac{v_{1}}{M_{1}}t_{cc}^{*}\left(k_{x}+ik_{y}\right)$.
The phase diagram at AB$_2$ stacking is shown in Fig. \ref{AB}.
The other parameters adopted in the numerical calculations are the same as
AA stacking and there are overall three phases found.

The general feature of the phase diagram resembles that of Fig. \ref{AA}(b).
At large band gap, the exciton density is extremely small and Coulomb
interaction has little effect on the NI phase. As the band gap is reduced
below the exciton binding energy, the system enters the QAH insulator phase
with spontaneous time reversal symmetry breaking, and finally turns
into the QSH insulator phase as the band gap is further reduced.
All these three phases are anisotropic in pairing gaps and quasiparticle
dispersions due to the anisotropy of interlayer coupling. However, the qualitative feature
of pairing gap $\Delta$ in each phase is the same as Fig. \ref{pd}.

\subsubsection{General translation}

\begin{figure}[h]
\centering \includegraphics[width=8cm]{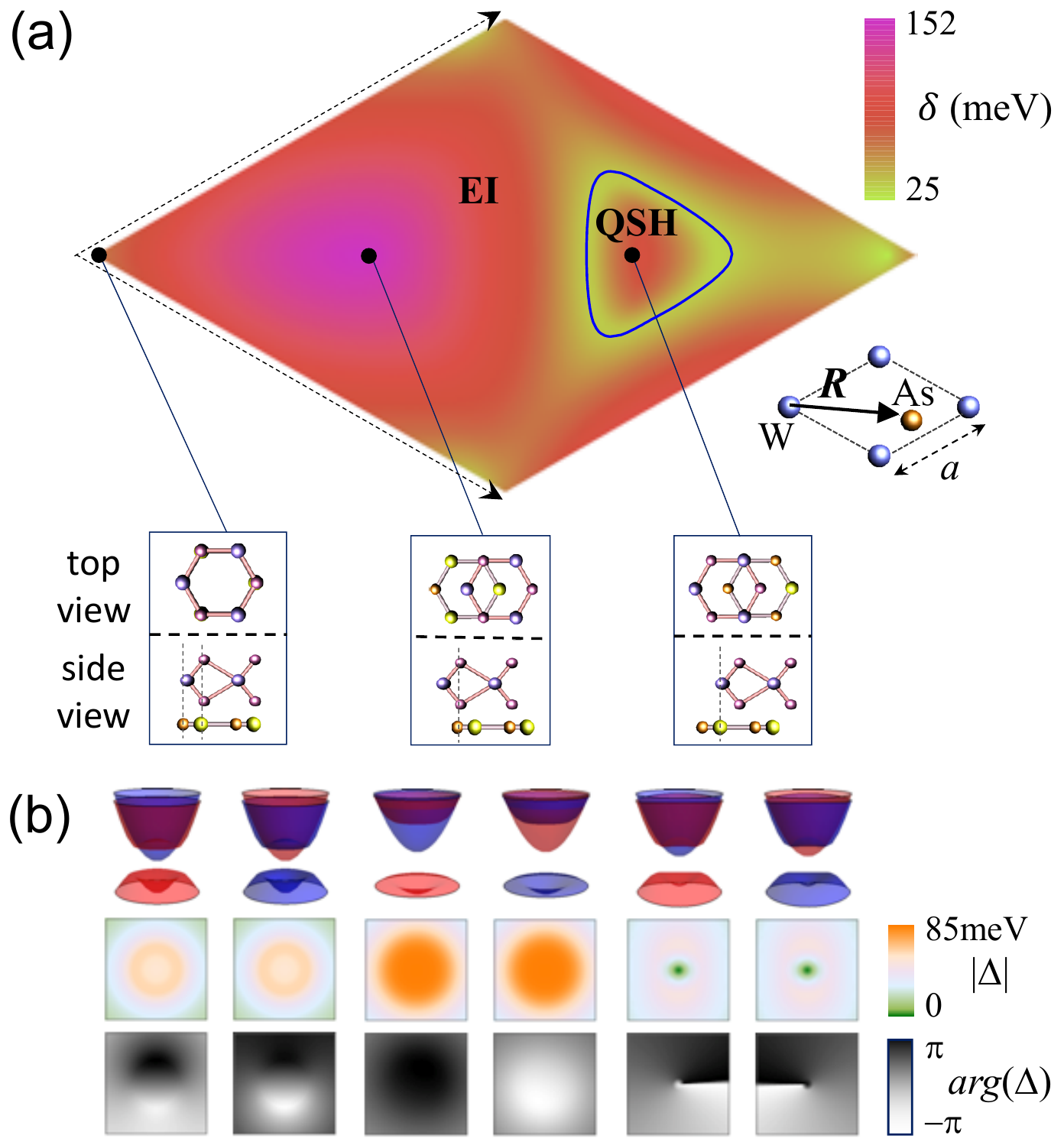} \caption{(Color online) (a) Phase diagram of AB stacking in a moir\'e supercell. The QSH/EI phase boundary is marked by blue line.
The nonzero gap at the boundary is also caused by finite discretization
interval of $\mathbf{k}$, similar to Fig. \ref{AAr}. $\epsilon_{\perp}=6$ and $E_g=-100$ meV.
 (b) Phases at three high symmetry locals,
along with top and side views of interlayer translation at these high
symmetry locals. The shown range of
momentum space is $\left[-\frac{\pi}{10a},\,\frac{\pi}{10a}\right]\times\left[-\frac{\pi}{10a},\,\frac{\pi}{10a}\right]$.}
\label{ABr} 
\end{figure}

Similarly, in the moir\'e supercell of AB stacking, the
phase diagram is shown in Fig. \ref{ABr}.
For regions close to the high symmetry local AB$_{2}$, the system lies in the QSH
insulator phase, with anisotropic pairing gap. Far away from the local AB$_{2}$, the
$s$-wave component of interlayer coupling dominates and the system
enters EI phase. At the boundary of this topological phase transition,
 band gap should close. Here also due to
finite discretization interval of $\mathbf{k}$-points in numerical
calculations, the minimal gap is nonzero.

\section{Discussion}

Type II band alignment forms between monolayer 
2H-MoTe$_{2}$/2H-WSe$_{2}$/2H-WTe$_{2}$ and hexagonal BAs, 
among which 2H-WSe$_{2}$ is the best choice
from the aspect of lattice match with BAs \cite{Band_alignment},
so that their heterobilayer consists of a moir\'e pattern with long period.
However, the main features predicted here also apply to other choices of TMD/BAs heterobilayer. The local approximation works well in predicting the local physics in moir\'e superlattices of long period, but is incapable of giving the moir\'e miniband structure.

At the high symmetry locals where the $s$-wave component of interlayer
tunneling vanishes, the phase of $s$-wave exciton condensate is unrestricted
and counterflow superfluidity is preserved. As an unique feature,
here the ES phase is also associated with an in-plane electric dipole \cite{zhu_gate_2019},
as the wave function of electron-hole relative motion is anisotropic
due to $s/p$-wave component mixing.
In other words, the overall electron-hole dipole tilts away from the out-of-plane direction.
 More specifically, analogous to cooper pair in 
Bardeen-Cooper-Schrieffer (BCS) theory of superconductivity \cite{BCS},
the exciton wave function
in momentum space is $F_{\mathbf{k}}=u_{\mathbf{k}}v_{\mathbf{k}}$,
Fourier transform of which gives the exciton wave function in real space
$F(\mathbf{r})$, with $\mathbf{r}$ being the in-plane component of electron-hole separation vector. 
The in-plane electric dipole per exciton is then
calculated as $\mathbf{d}_{ex}=e\int d\mathbf{r}F(\mathbf{r})^{*}\mathbf{r}F(\mathbf{r})/\int d\mathbf{r}|F(\mathbf{r})|^{2}$.
Decomposing the pairing gap function in terms of $s$- and $p$-wave components,
$\Delta_{\tau\mathbf{k}}=\tau\Delta_{s}(\mathbf{k})e^{-i\tau\theta}-\tau\Delta_{p}(\mathbf{k})^{-i\tau\phi(\mathbf{k})}$,
with $\phi(\mathbf{k})$ being the azimuth angle of $\mathbf{k}$
and $\theta$ being the unrestricted condensate phase, the orientation
of the in-plane electric dipole is uniquely determined by the condensate
phase. {\it In situ} measurement of the in-plane electric dipole can
therefore give direct information about the local condensate phase
$\theta$. Note that only for pairing gap with both $s$- and $p$-wave
components, the in-plane electric dipole is nonzero, while in other cases with $s$-
or $p$-wave component alone, the in-plane electric dipole always
vanishes and the overall electric dipole points perpendicular to the layer plane.

In the moir\'e supercell of AA stacking, the phase at high symmetry
local AA$_{0}$ is an ES, as $s$-wave component of interlayer
tunneling vanishes there. Away from AA$_{0}$, the strength
of $s$-wave tunneling gradually increases, which tends to pin the phase of
exciton condensate and therefore destroys counterflow superfluidity. So
there is a crossover from ES phase to EI phase as the spatial deviation
from AA$_{0}$ stacking region increases. The moir\'e superlattice consists of patterns
of ES and QSH phases separated by EI phases. Quantitative features
of the ES to EI phase crossover can be explored using a miminal-field-theory
model \cite{moon_spontaneous_1995,su_how_2008}, and will be an interesting subject
for future study.

There is numerical evidence of gap closing at the NI and QAH phase
boundary in Figs. \ref{AA}(b) and \ref{AB}, though due to symmetry
change across the topological phase transition, the gap does not necessarily
close in principle \cite{ezawa_topological_2013}. The gap-non-closing features
at ES/QSH and QAH/QSH phase boundaries occur as first-order quantum
phase transitions, like those reported in other systems \cite{zhu_gate_2019}.

The phase diagrams in Figs. 4 and 6 demonstrate the versatility of phases tunable experimentally by interlayer translation, together with the gate switchability of these phases, making this van der Waals heterobilayer advantageous over previous proposals of large-gap quantum spin Hall insulator systems.

\section{Conclusion}

Using WSe$_2$/BAs heterobilayer as an example,
we show that van der Waals heterobilayer of TMD/BAs can be a candidate system for realizing
a QSH insulator with sizable band gap, which can be tuned between the topological and trivial phases by both interlayer bias and interlayer translation. 
This heterobilayer features a type
II band alignment, with spin doubly degenerate conduction band and spin-valley locked
valence band from BAs and TMD layer, respectively. When the bands get inverted by an interlayer bias, hybridization leads to a stacking-configuration dependent topological band
inversion. However, in the absence of
Coulomb interaction, the double spin degeneracy of BAs conduction band leaves an
 un-hybridized conduction
band inside the hybridization gap, and hence the system is a spin-valley locked metal. Further taking into account the effect of Coulomb interaction, we find the
un-hybridized conduction band can be shifted upwards in energy due to the interaction with the
hybridization induced electric dipole associated with charge transfer between the two layers. 
A sizable global gap, of the order of hundred meV, appears
in a broad parameter regime, tunable by both interlayer bias and stacking registry.
Besides, the helical edge states in the QSH insulator phases
can be switched on/off by the interlayer bias alone.
Therefore, this heterobilayer provides a promising
platform for engineering large gap QSH insulator and its moir\'e
pattern in electrically tunable ways.

\begin{acknowledgements} The work is mainly supported by the Croucher Foundation
(Croucher Innovation Award), the RGC of HKSAR (HKU17302617, C7036-17WF). Yong Wang acknowledges support by NSFC Project Nos. 11604162 and 61674083. \end{acknowledgements}

\appendix{} 

\section{Three-band model without interlayer hopping}

The ground state wave function of the three-band model is $|\Psi\rangle=\prod_{\tau\mathbf{k}}(u_{\tau\mathbf{k}}+v_{\tau\mathbf{k}s}\hat{a}_{\mathbf{k}s}^{\dagger}\hat{b}_{\mathbf{k}s}+v_{\tau\mathbf{k}\bar{s}}\hat{a}_{\mathbf{k}\bar{s}}^{\dagger}\hat{b}_{\mathbf{k}s})|0\rangle$,
where $s=\uparrow\left(\downarrow\right)$, for $\tau=1\left(-1\right)$, 
and $\bar{s}$ reverses the value of $s$. The mean-field energy reads
\begin{align}
\langle H\rangle & =\sum_{\tau\mathbf{k}}\left(\varepsilon_{\mathbf{k}cs}-\varepsilon_{\mathbf{k}vs}\right)\left(|v_{\tau\mathbf{k}s}|^{2}+|v_{\tau\mathbf{k}\bar{s}}|^{2}\right)\nonumber \\
 & -\frac{1}{2}\sum_{\tau\mathbf{k},\mathbf{k}'s}U(\mathbf{k}-\mathbf{k}')|v_{\tau\mathbf{k}s}|^{2}|v_{\tau\mathbf{k}'s}|^{2}\nonumber \\
 & -\sum_{\tau\mathbf{k},\mathbf{k}'}U(\mathbf{k}-\mathbf{k}')v_{\tau\mathbf{k}s}v_{\tau\mathbf{k}\bar{s}}^{*}v_{\tau\mathbf{k}'s}^{*}v_{\tau\mathbf{k}'\bar{s}}\nonumber \\
 & -\sum_{\tau\mathbf{k},\mathbf{k}'s'}V(\mathbf{k}-\mathbf{k}')u_{\tau\mathbf{k}s}v_{\tau\mathbf{k}s'}^{*}u_{\tau\mathbf{k}'s}^{*}v_{\tau\mathbf{k}'s'}\nonumber \\
 & -\frac{1}{2}\sum_{\tau\mathbf{k},\mathbf{k}'}U(\mathbf{k}-\mathbf{k}')\sum_{s}|v_{\tau\mathbf{k}s}|^{2}\sum_{s}|v_{\tau\mathbf{k}'s}|^{2}\nonumber \\
 & +\frac{2\pi e^{2}d}{\epsilon}\left(\sum_{\tau\mathbf{k}s}|v_{\tau\mathbf{k}s}|^{2}\right)^{2}.
\end{align}
Here the summation over $\tau$ alone means the value of $s$
is simultaneously determined by the relation $s=\uparrow\left(\downarrow\right)$,
for $\tau=1\left(-1\right)$,
 while the summation over both
$\tau$ and $s$ indicates their values are independent.

Parameterize the mean-field wave function as $u_{\tau\mathbf{k}s}=\cos\theta_{\tau\mathbf{k}}$,
$v_{\tau\mathbf{k}s}=\sin\theta_{\tau\mathbf{k}}\cos\phi_{\tau\mathbf{k}}$,
$v_{\tau\mathbf{k}\bar{s}}=\sin\theta_{\tau\mathbf{k}}\sin\phi_{\tau\mathbf{k}}e^{i\varphi_{\tau\mathbf{k}}}$
and then the normalization condition $|u_{\tau\mathbf{k}s}|^{2}+|v_{\tau\mathbf{k}s}|^{2}+|v_{\tau\mathbf{k}\bar{s}}|^{2}=1$
is readily satisfied. The above mean-field energy can be expressed
as 
\begin{align}
\langle H\rangle & =\sum_{\tau\mathbf{k}}\left(\varepsilon_{\mathbf{k}c}-\varepsilon_{\mathbf{k}v}\right)\sin^{2}\theta_{\tau\mathbf{k}}\nonumber \\
 & -\frac{1}{2}\sum_{\tau\mathbf{k},\mathbf{k}'}U(\mathbf{k}-\mathbf{k}')\sin^{2}\theta_{\tau\mathbf{k}}\sin^{2}\theta_{\tau\mathbf{k}'}\nonumber \\
 & \times\left[\cos^{2}\phi_{\tau\mathbf{k}}\cos^{2}\phi_{\tau\mathbf{k}'}+\sin^{2}\phi_{\tau\mathbf{k}}\sin^{2}\phi_{\tau\mathbf{k}'}\right.\label{eq:phi1-1}\\
 & \left.+2\cos\phi_{\tau\mathbf{k}}\sin\phi_{\tau\mathbf{k}}\cos\phi_{\tau\mathbf{k}'}\sin\phi_{\tau\mathbf{k}'}e^{i(\varphi_{\tau\mathbf{k}'}-\varphi_{\tau\mathbf{k}})}\right]\label{eq:phi2-1}\\
 & -\sum_{\tau\mathbf{k},\mathbf{k}'}V(\mathbf{k}-\mathbf{k}')\cos\theta_{\tau\mathbf{k}}\sin\theta_{\tau\mathbf{k}}\cos\theta_{\tau\mathbf{k}'}\sin\theta_{\tau\mathbf{k}'}\nonumber \\
 & \times\left[\cos\phi_{\tau\mathbf{k}}\cos\phi_{\tau\mathbf{k}'}+\sin\phi_{\tau\mathbf{k}}\sin\phi_{\tau\mathbf{k}'}e^{i(\varphi_{\tau\mathbf{k}'}-\varphi_{\tau\mathbf{k}})}\right]\label{eq:phi3-1}\\
 & -\frac{1}{2}\sum_{\tau\mathbf{k},\mathbf{k}'}U(\mathbf{k}-\mathbf{k}')\sin^{2}\theta_{\tau\mathbf{k}}\sin^{2}\theta_{\tau\mathbf{k}'}\nonumber \\
 & +\frac{2\pi e^{2}d}{\epsilon}\left(\sum_{\tau\mathbf{k}}\sin^{2}\theta_{\tau\mathbf{k}}\right)^{2}.
\end{align}
It can be easily seen that minimization of the expressions inside
the square brackets in lines \ref{eq:phi1-1}-\ref{eq:phi3-1} with
respect to $\phi_{\mathbf{k}}$ and $\varphi_{\mathbf{k}}$ leads to the condition that
these two phase angles are both constants independent of $\mathbf{k}$.
The general $3\times3$ mean-field Hamiltonian in the
basis $(\hat{a}_{\mathbf{k}\bar{s}},\hat{a}_{\mathbf{k}s},\hat{b}_{\mathbf{k}s})$
are $\hat{H}_{\tau}=\sum_{\mathbf{k}}\left(\hat{a}_{\mathbf{k}\bar{s}}^{\dagger},\hat{a}_{\mathbf{k}s}^{\dagger},\hat{b}_{\mathbf{k}s}^{\dagger}\right)h_{\tau\mathbf{k}}\left(\hat{a}_{\mathbf{k}\bar{s}},\hat{a}_{\mathbf{k}s},\hat{b}_{\mathbf{k}s}\right)^{T}$, and
\begin{align}
h_{\tau\mathbf{k}} & =\nonumber \\
 & \begin{bmatrix}\varepsilon_{\mathbf{k}c}-\xi_{\tau\mathbf{k}a\bar{s}}+\xi_0 & -\chi_{\tau\bar{s}s}(\mathbf{k}) & -\Delta_{\tau\bar{s}s}(\mathbf{k})\\
-\chi_{\tau\bar{s}s}^{*}(\mathbf{k}) & \varepsilon_{\mathbf{k}c}-\xi_{\tau\mathbf{k}as}+\xi_0 & -\Delta_{\tau ss}(\mathbf{k})\\
-\Delta_{\tau\bar{s}s}^{*}(\mathbf{k}) & -\Delta_{\tau ss}^{*}(\mathbf{k}) & \varepsilon_{\mathbf{k}v}+\xi_{\tau\mathbf{k}bs}-\xi_0
\end{bmatrix},
\end{align}
where $\xi_0=\frac{2\pi e^{2}d}{\epsilon}\sum_{\tau\mathbf{k}s}|v_{\tau\mathbf{k}s}|^{2}$,
$\xi_{\tau\mathbf{k}bs}=\sum_{\mathbf{k}'}U(\mathbf{k}-\mathbf{k}')\left(|v_{\tau\mathbf{k}'s}|^{2}+|v_{\tau\mathbf{k}'\bar{s}}|^{2}\right)$, and
$\xi_{\tau\mathbf{k}as}=\sum_{\mathbf{k}'}U(\mathbf{k}-\mathbf{k}')|v_{\tau\mathbf{k}'s}|^{2}$.
The pairing gaps are defined as $\Delta_{\tau s's}(\mathbf{k})=\sum_{\mathbf{k}'}V(\mathbf{k}-\mathbf{k}')u_{\tau\mathbf{k}'s}^{*}v_{\tau\mathbf{k}'s'}$,
$\chi_{\tau\bar{s}s}(\mathbf{k})=\sum_{\mathbf{k}'}U(\mathbf{k}-\mathbf{k}')v_{\tau\mathbf{k}'s}^{*}v_{\tau\mathbf{k}'\bar{s}}$. With the condition that $\phi_{\mathbf{k}}$ and $\varphi_{\mathbf{k}}$ are constants, namely, $v_{\tau\mathbf{k}s}/v_{\tau\mathbf{k}\bar{s}}=
\mathrm{constant}$,
the general Hamiltonian can be diagonalized to give three quasiparticle band dispersions, one
of which is decoupled from the other two. The two coupled bands have
dispersion relations $E_{\tau\mathbf{k}\pm}=\frac{\varepsilon_{\mathbf{k}c}+\varepsilon_{\mathbf{k}v}}{2}\pm\sqrt{\left(\frac{\varepsilon_{\mathbf{k}c}-\varepsilon_{\mathbf{k}v}}{2}-\xi_{\tau\mathbf{k}bs}+\xi_0\right)^{2}+|\Delta_{\tau ss}(\mathbf{k})|^{2}}$,
and the decoupled one is $E_{\mathbf{k}c}=\varepsilon_{\mathbf{k}c}+\xi_0$.

\section{First-principles calculation results}

\begin{figure}[h]
\centering \includegraphics[width=\linewidth]{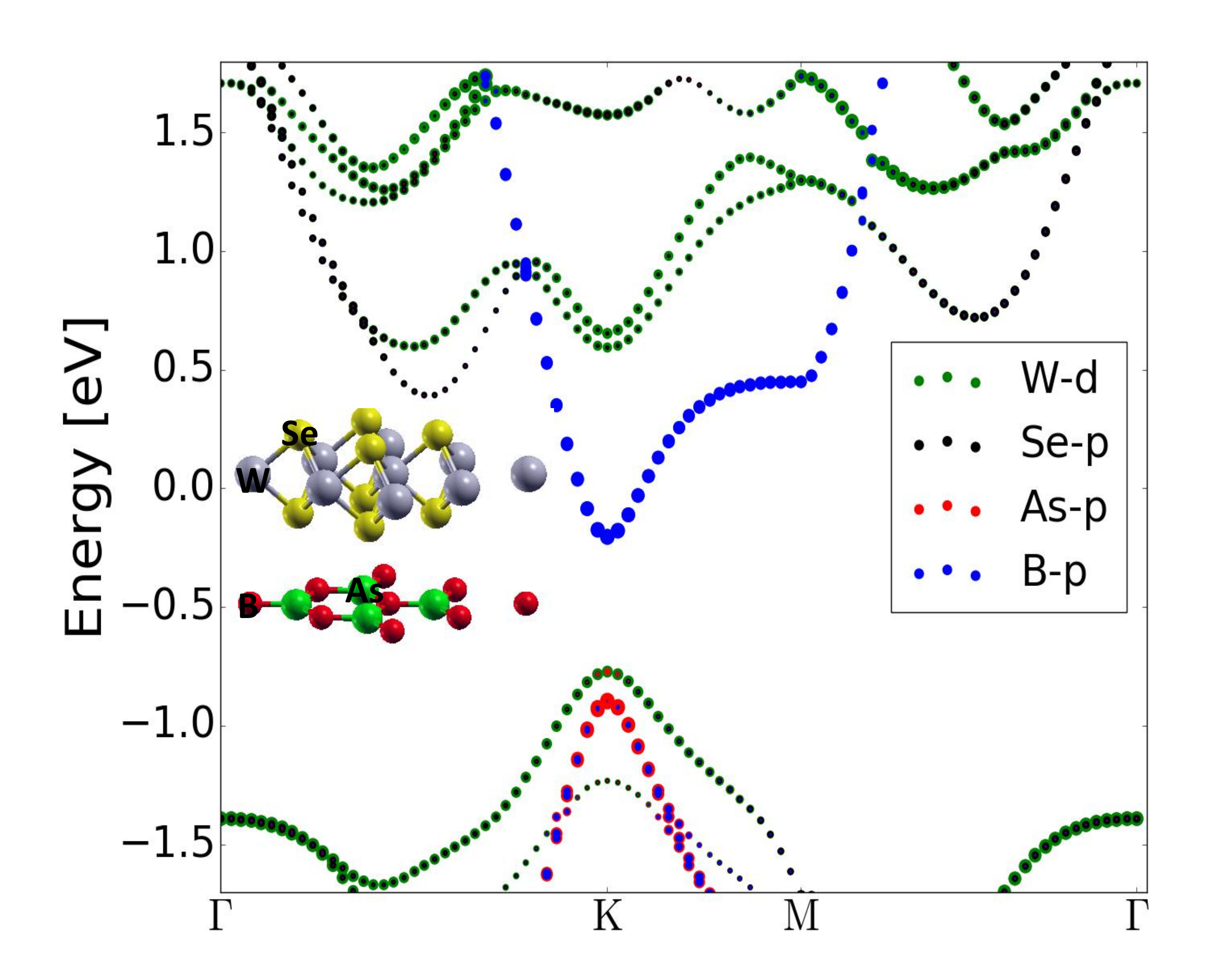}
\caption{(Color online) Orbital projected band structure of a AA-type WSe$_{2}$/BAs heterobilayer. Inset: crystal structure of the heterobilayer.}
\label{FigS1} 
\end{figure}

The band structure of a AA-type WSe$_{2}$/BAs heterobilayer, as demonstrated in Fig.~\ref{FigS1}, is calculated with density functional theory implemented in {\sc Quantum Espresso} software \cite{QE-2009,QE-2017}.  The full-relativistic pseudo-potential with local density approximation (LDA) for the exchange-correlation functional has been used to include the spin-orbital coupling. The lattice constant of the WSe$_{2}$/BAs heterobilayer is fixed as $3.30$~\AA, and its crystal structure is optimized with the Broyden-Fletcher-Goldfarb-Shanno (BFGS) quasi-Newton algorithm. The convergence thresholds on the total energy and force are set as $10^{-4}$~Ry and $10^{-3}$~Ry/Bohr, respectively. The kinetic energy cutoff for the plane-wave function is set as $80$~Ry, and the convergence threshold for the electronic self-consistent calculation is set as $10^{-10}$~Ry. The $\mathbf{k}$-points in Brillouin zone are sampled by the 15$\times$15$\times$1 Monkhorst-Pack grid.  With the correction of van der Waals interaction by the vdW-DF method, the layer distance between the W and B atoms is determined as $5.7$~\AA. The band structure in Fig.~\ref{FigS1} shows that the WSe$_{2}$/BAs heterostructure has a type II band alignment, where the valence band comes from the WSe$_{2}$ layer with a large spin splitting and the conduction band comes from the BAs layer with negligible spin splitting. The direct band gap at $K$ point for this structure is about $0.568$~eV. Similar band structures have also been obtained for the heterobilayer with other stacking registries.

\begin{table}[htb]
\centering %
\noindent\begin{minipage}[t]{1\linewidth}%
\caption{The optimized layer distance $D$, the valence band splitting $\lambda_{v}$ and the conduction band splitting $\lambda_{c}$ at the $K$ point for bilayer BAs with different stacking manners. The lattice constant is fixed as $3.39$~\AA.}
\begin{tabular}{c|c|c|c|c|c|c}
\hline 
\hline
              &  AA$_{0}$  & AA$_{1}$  & AA$_{2}$  & AB$_{0}$ & AB$_{1}$ & AB$_{2}$ \\
\hline 
\hline
$D$ (\AA)            &  4.50 & 4.00  & 4.00   & 4.42 &  3.97 &  4.29 \\
\hline 
$\lambda_{v}$ (meV) &  442  &  187  &  188   &  1   &  1    &  588 \\
\hline 
$\lambda_{c}$ (meV) &  340  &  230  &  230   &  1   &  688  &  1   \\
\hline 
\hline
\end{tabular}\label{tableII}%
\end{minipage}
\end{table}

The band structures of bilayer BAs with different stacking registries have also been calculated in order to estimate the interlayer coupling $t_{ij}^{\mathrm{BAs}}$. Here, the Perdew-Burke-Ernzerhof exchange-correlation functional without spin-orbital coupling has been exploited. By fixing the lattice constant as $3.39$~\AA, the optimized layer distances and the values of band splitting at the valence band maximum and conduction band minimum are listed in Table~\ref{tableII}.

\bibliographystyle{iopart-num}
%\bibliography{reference}

\end{document}